\documentclass[fleqn,10pt]{wlscirep}

\usepackage{bm}

\title{Radiation emission in laser-wakefields driven by structured laser pulses with orbital angular momentum}

\author[1,*]{Joana Lu\'is Martins}
\author[2]{Jorge Vieira}
\author[1]{Julien Ferri}
\author[1]{T\"unde F\"ul\"op}
\affil[1]{Chalmers University of Technology, Department of Physics, Gothenburg, 41296, Sweden}
\affil[2]{GoLP/Instituto de Plasmas e Fus\~ao Nuclear, Instituto Superior T\'ecnico, Universidade de Lisboa, Lisboa, 1049-001, Portugal}

\affil[*]{ljoana@chalmers.se}

\begin{abstract}
High-intensity X-ray sources are invaluable tools, enabling experiments at the forefront of our understanding of materials science, chemistry, biology, and physics.
Laser-plasma electron accelerators are sources of high-intensity X-rays, as electrons accelerated in wakefields emit short-wavelength radiation due to betatron oscillations. 
While applications such as phase-contrast imaging with these betatron sources have already been demonstrated, others would require higher photon number and would benefit from increased tunability.
In this paper we demonstrate, through detailed 3D simulations, a novel configuration for a laser-wakefield betatron source that increases the energy of the X-ray emission and also provides increased flexibility in the tuning of the X-ray photon energy.
This is made by combining two Laguerre-Gaussian pulses with non-zero net orbital angular momentum, leading to a rotation of the intensity pattern, and hence, of the driven wakefields. The helical motion driven by the laser rotation is found to dominate the radiation emission, rather than the betatron oscillations. Moreover, the radius of this helical motion can be controlled through the laser spot size and orbital angular momentum indexes, meaning that the radiation can be tuned fully independently of the plasma parameters.\end{abstract}
\begin{document}

\flushbottom
\maketitle
\thispagestyle{empty}

\section*{Introduction}

High brightness X-ray sources are of great use throughout science, from diagnosing biological samples~\cite{Cole_SciRep_2004} to probing extreme states of matter~\cite{Albert_PPCF_2016,Albert_ProcSpie_2017}. Conventional methods for the production of X-rays require first accelerating electrons to high energies, which involves large and costly facilities such as synchrotrons. Laser-driven accelerators \cite{TajimaDawson_PRL_1979} are potentially a much more compact and less costly approach to generate the required electrons and, for some electron energies, the X-ray emission as well ~\cite{Albert_PPCF_2016}.

The achievable electron energy in laser-wakefield accelerators (LWFAs) has undergone a rapid growth since the original  ``dream beam'' experiments \cite{Geddes_Nature_2004,Mangles_Nature_2004,Faure_Nature_2004}, in which mono-energetic multi-MeV electron beams were first obtained. Current wakefield accelerators now reach up to GeV energies with only a few centimetres of plasma as the acceleration medium \cite{Leemans_NatPhys_2006,Clayton_PRL_2010,Leemans_PRL_2014,Hyung_Taek_SciRep_2017}. 

Plasma wakefield accelerators not only provide high energy beams but represent a source of short-wavelength radiation as well. The transverse electric field that is naturally formed by the positive charge in the wakefield bubble causes the electrons to oscillate as they are being accelerated, which leads to the emission of betatron radiation~\cite{Whittum_POF_1992,Esarey_PRE_2002}. Due to the high electron energies attained in the accelerator, this emission is in the X-ray range, with typical peak photon energies up to 1-20 keV~\cite{Rousse_PRL_2004,TaPhuoc_PoP_2005,Kneip_2010}. This has been shown to be a viable broadband high-brightness X-ray source~\cite{Rousse_PRL_2004,Kneip_2010,Corde_RevModPhys_2013}.

The specific properties of the emission depend on the plasma density, electron energy and amplitude of electron oscillations. For stable laser propagation and maximum energy transfer to the electrons, it is desirable to operate under the so-called ``matching'' conditions \cite{Lu_PRSTAB_2016}. This fixes the plasma density as a function of the laser parameters, thus restricting the available parameter space. Several approaches have been explored to control the amplitude of the transverse oscillations. One possibility lies in modifying the initial amplitude of oscillation by controlling the phase-space of the injected electrons, for example through colliding pulse injection \cite{Corde_PRL_2011} or magnetically controlled injection \cite{Vieira_PPCF_2012}.

Alternative schemes have been proposed which modify not just the initial conditions but also the oscillation amplitude throughout the propagation of the wakefield. These include, for example, injection of the laser driver off-axis \cite{Rykovanov_PRL_2015} and at an angle with the plasma channel \cite{Luo_SciRep_2016}. These designs require good control of the laser incidence on the plasma channel. Laser drivers with duration close to the bubble size can be used to induce direct laser acceleration \cite{Nemeth_PRL_2008} and, for longer pulses (with ps duration), self-modulated laser-wakefield acceleration has been explored as an X-ray source \cite{Albert_PRL_2017,Ferri_PRSTAB_2016} for high-density matter probing. In all of these configurations, the laser driver has a Gaussian profile, which is also the typical profile used in the vast majority of experiments in this area. Though there has been much progress, some applications of X-ray sources are still out of reach due to insufficient number of photons and would benefit greatly from increased tunability \cite{Albert_PPCF_2016}.

The recent development of high-intensity light pulses with orbital angular momentum (OAM)~\cite{Brabetz_PoP_2015}
(for a theoretical overview see e.g.~\cite{Padgett_OptExpress_2017,Yao_AdvOptPhoton_2011}), such as Laguerre-Gaussian~\cite{Allen_PRA_1992} pulses, opens the way for new avenues to control laser-wakefield accelerators. Laguerre-Gaussian laser pulses have been shown to drive wakefields whose field structure is capable of accelerating positrons \cite{Vieira_PRL_2014}. Combinations of such beams can be used to change the topology of the wakefield itself, e.g. helical wakefields driven by ``light springs'' \cite{Jorge_arXiv_2018}.  Superposed Laguerre-Gaussian lasers have also been used, in the low-amplitude (linear) regime, to produce carefully-tailored intensity patterns~\cite{Cormier_Michel_PhysRevSTAB_2011}. This setup holds particular promise for developing laser-wakefield based free-electron lasers~\cite{Wang_SciRep_2017}.

In this work, we explore the use of such composite vortex beams to instead enhance and control incoherent X-ray emission. We propose a novel helical wiggler radiation source based on a pair of nonlinear rotating wakefields. These are driven by a laser composed of two ultra-short ultra-intense Laguerre-Gaussian beams with non-zero net Orbital Angular Momentum (OAM) index. In this proposed design, the combination of beams is employed to induce a rotation of the wakefields. This rotation forces the helical motion of the trapped electrons, enhancing their oscillation amplitude. This oscillation saturates with its amplitude approximately equal to the radius of the helix described by the laser lobes, which is given by the distance to the propagation axis and is larger than the bubble radius. Tuning of the oscillation amplitude (and hence of the wiggler strength parameter) is achieved by controlling this distance, which can be done by varying the laser spot size and the orbital angular momentum indexes. This permits greater control over the radiation energy and spectrum compared to typical LWFA betatron sources.  So long as blowout occurs, the wiggler strength in our scheme does not depend sensitively on the laser intensity unlike~\cite{Wang_SciRep_2017} or direct laser acceleration. In addition, this pulse configuration is compatible with many methods of electron injection into the wakefields, providing tuning capability regardless of the injection mechanism.

X-ray generation in this scheme is also drastically enhanced.
We show, through three-dimensional particle-in-cell (PIC) simulations, that a ten-fold increase of the radiated energy per electron is possible compared to non-rotating wakefields driven by lasers with the same energy and duration, and similar waist. Finally, we observe that the field structure of these wakefields is itself different from both the regime studied in \cite{Wang_SciRep_2017} and that of typical Gaussian-pulse-driven wakefields in the nonlinear regime, with the appearance of an additional magnetic field component in the laser propagation direction.

\section*{Results}

In this work, we investigate radiation emission from laser-wakefield accelerators (LWFAs) driven by multiple Laguerre-Gaussian (LG) pulses. These pulses are characterised by an electric field of the form
\begin{eqnarray}
E_{\mathrm{laser}} \propto  \left( \frac{\rho}{W_0} \right)^{|\ell|} \mathrm{L}_p^{|\ell|}\left( \frac{2 \rho^2}{W_0^2} \right) \exp  \left( -\frac{\rho^2}{W_0^2} \right) \times \cos \left[ \omega_0 t - k_0z + \psi (z) + \frac{1}{2}\rho^2 \frac{k}{R(z)} + \ell\theta \right],
\end{eqnarray}
in cylindrical coordinates, where $\mathrm{L}_p^{|\ell|}$ designates the Laguerre-Gaussian generalised polynomial with radial index $p$ and azimuthal orbital angular momentum (OAM) index $\ell$, $\rho$ is the radial distance to the propagation axis, $W_0$ is the laser spot size, $\omega_0$ is the laser central frequency, $k_0$ is its wavenumber, $z$ is the coordinate along the laser propagation axis, $\psi (z) = (2 p + \ell + 1) \arctan (z/z_R)$ is the Gouy shift, $z_R$ is the Rayleigh length and $R(z)$ is the curvature, given by $R(z) = z \left[ 1 + ( z/z_R)\right]$.

While individual Laguerre-Gaussian beams have annular intensity profiles, more complex patterns can emerge from the combination of multiple beams. In particular, combinations of LG pulses with zero radial index lead to composite vortex beams \cite{Yao_AdvOptPhot_2011}. Examples of these composite vortex beams are ``light springs'', a combination of pulses with a helical intensity pattern \cite{Pariente_OptLett_2015}. It has been shown that these light springs can drive twisted helical wakefields, which rotate as the driver beam propagates \cite{Jorge_arXiv_2018}. If the duration of such pulses is sufficiently short, they will no longer look like a spring, instead being comprised of multiple rotating laser beam lobes, each driving its own wakefield. The number of lobes, their size and their azimuthal position are then determined by the OAM indexes $\ell$, the amplitudes, the spot sizes and frequencies of the pulses involved. It is this regime that we explore in this work.

The proposed setup consists of a composite vortex beam, comprised of two LG laser pulses, that is sent through an underdense plasma. The two pulses have frequencies $\omega_{1,2}$, wavenumbers $k_{1,2}$, spot sizes $W_{0\,(1,2)}$, and amplitudes $E_{1,2}$. Hence, the combined beam has a profile (in vacuum) of the form 
\begin{eqnarray}
I_{\mathrm{laser}} \propto E_1^2 + E_2^2 + 2 E_1 E_2 \cos \left[ \Delta\omega\, t - \Delta k\, z + \Delta \ell + \Delta\theta - \Delta \psi (z) + \frac{1}{2}\rho^2 \left( \frac{k_2}{R_2(z)}-\frac{k_1}{R_1(z)} \right) \right], 
\end{eqnarray}
in cylindrical coordinates. Here, $\Delta\omega = \omega_1 - \omega_2$, $\Delta k = k_1 - k_2$, $\Delta \ell = \ell_1 - \ell_2$, and $\Delta \psi (z) = \psi_1(z) - \psi_2(z)$. The first terms are dependent only on radius and the cosine term is responsible for the intensity pattern composed of $\ell_2 - \ell_1$ lobes distributed azimuthally \cite{Galvez_ProcSPIE_2006}. If  $\ell_2 \ne - \ell_1$ and either ${k_0}_{(1)} \ne {k_0}_{(2)}$ or the pulse is not cylindrically symmetric, the lobes of a composite vortex beam will rotate as the beam propagates in vacuum and crosses its focal plane \cite{Baumann_OptExp_2009,Galvez_ProcSPIE_2006}. Recently, Vieira \textit{et al} have shown that, in the linear regime, the same rotation will occur for light springs propagating in an underdense plasma \cite{Jorge_arXiv_2018,Vieira_APS_2017}. 

To further explore this setup and compare with a typical Gaussian-pulse LWFA, detailed three-dimensional particle-in-cell (PIC) simulations were performed within the OSIRIS framework \cite{osiris1,osiris2}. The laser parameters were chosen to keep the laser energy constant in all the simulations. In the first two simulations, the laser driver is composed of two LG pulses with the same normalised peak vector potential and pulse length, and approximately the same frequency. The spot size is of the same order but was adjusted so that in all the cases the laser energy is the same. The LG radial index is $p=0$ in all cases and the azimuthal indexes are $\ell_1 = 3$ and $\ell_2 = 5$ in simulation A, and $\ell_1 = 1$ and $\ell_2 = -1$ in simulation B. In the third simulation, which will be referred to as simulation C, the driver is a standard Gaussian laser pulse with the same frequency and pulse length. 

In each case, the laser driver propagates in a plasma channel with a transverse parabolic density profile matched to its spot size, given by $n(r) = n_0 \left(1 + r_0^2\,\Delta n / n_0\right)$, where $r_0 = k_p W_0$ is the channel radius, $k_p$ is the electron plasma wavenumber and $\Delta n / n_0 = 4/(k_p^4 W_0^4 )$. A mixture of Hydrogen and a small percentage of Nitrogen is used to achieve ionisation injection \cite{Pak_PRL_2010,McGuffey_PRL_2010}. This avoids the need for external injection, which would require fine control of the initial beam parameters. Further simulation details are described in the Methods section.

\subsection*{Dynamics of wakefields}

\begin{figure}[ht]
\centering
\includegraphics[width=\linewidth]{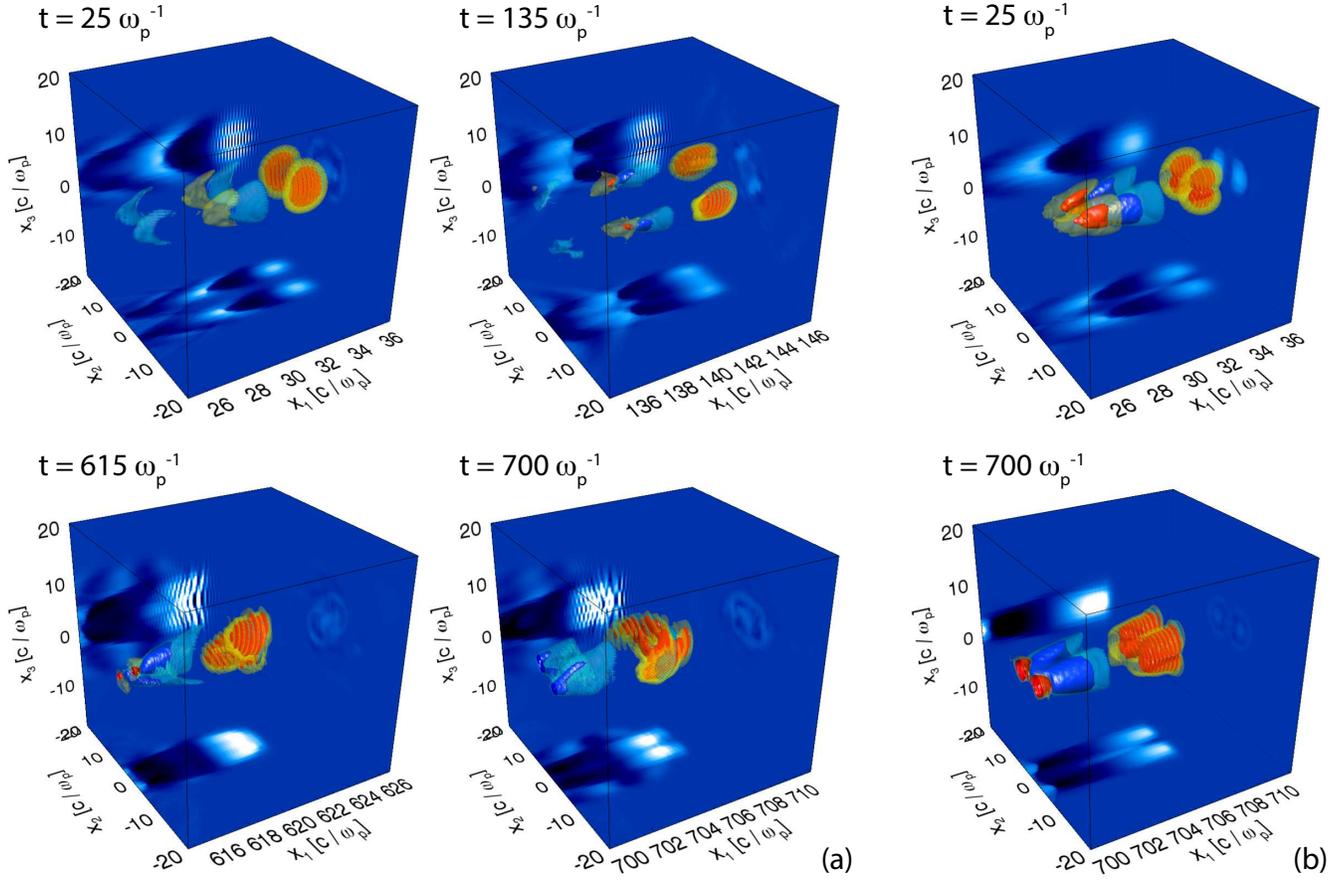}
\caption{Isosurfaces of the electric field component in the laser propagation direction at times $t \simeq 25 \,\omega_p^{-1}$, $135 \,\omega_p^{-1}$, $615 \,\omega_p^{-1}$ and $700 \,\omega_p^{-1}$ for simulation A (left panels) and $t \simeq 25 \,\omega_p^{-1}$ and $700 \,\omega_p^{-1}$ for simulation B (right panels). Two distinct wakefields are observed in each case, which rotate over time in simulation A but not in B.}
\label{fig:e1_rot_non_rot_wake}
\end{figure}

In Figure \ref{fig:e1_rot_non_rot_wake}(a), the longitudinal component of the electric field is depicted for simulation A ($\ell_1 = 3$, $\ell_2 = 5$) for four different times. As the laser intensity lobes propagate and rotate in the underdense plasma (for $\ell_2 \ne - \ell_1$), they will drive wakefields behind them. If the rotation frequency, $\Omega_h = 2 c^2 / (\omega_0 W_0^2)\, \mathrm{sign (\ell_0)}$ (from linear theory \cite{Jorge_arXiv_2018}), is much smaller than the electron plasma frequency $\omega_p$, the bubbles will be able to follow the laser driver rotation. For our configuration, this condition becomes
\begin{equation}
\frac{2 \omega_p}{\omega_0}\frac{1}{W_0^2 k_p^2} \ll 1.
\end{equation}
 Operation  in  the  matched  regime  typically  favours  lower densities, which means this condition is easily fulfilled, as in this simulation. 

From our analysis of the evolution of the laser fields in the simulation, the rotation frequency was determined to be approximately $\Omega_h \sim 9\times 10^{-3} \,\omega_p$, which is similar to the theoretically predicted value of $\Omega_h \simeq 7\times 10^{-3} \,\omega_p$. The discrepancy may be due to laser depletion during the propagation, which is not accounted for in the theoretical estimate. For a zero net orbital angular momentum, as in simulation B ($\ell_1 = 1$, $\ell_2 = -1$), no wakefield rotation is observed (Figure \ref{fig:e1_rot_non_rot_wake}(b)), as expected from theory. 
\begin{figure}[ht]
\centering
\includegraphics[width=\linewidth]{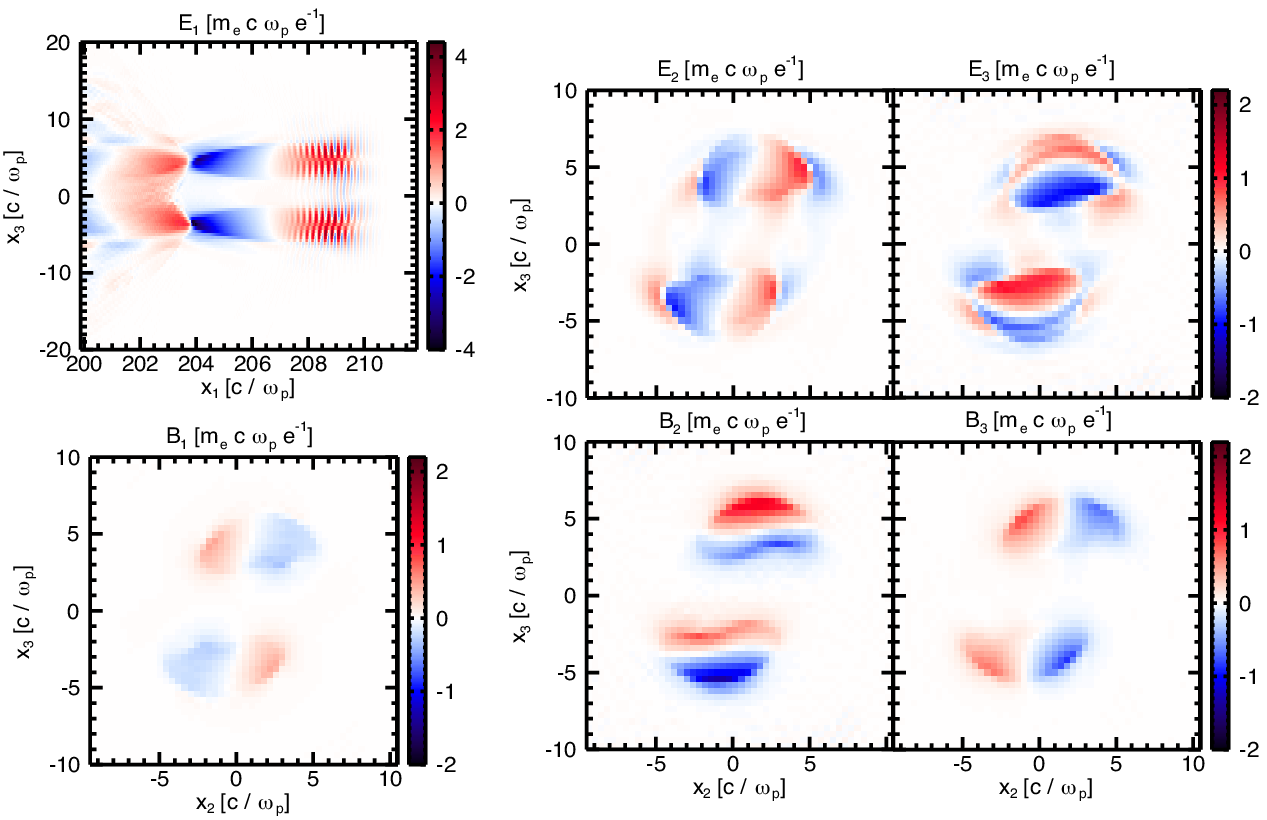}
\caption{Electric and magnetic fields inside the two rotating bubbles at $t = 199.88 \, {\omega_p}^{-1}$. The longitudinal electric field $E_1$ is a slice at $x_3 = 0\, c/\omega_p$ (a). All other field components: $E_2$ (b), $E_3$ (c), $B_1$ (d), $B_2$ (e) and $B_3$ (f) are slices taken at $x_1 = 205.3\, c/\omega_p$.}
\label{fig:flds_rot_wake_bubbles}
\end{figure}

Before delving into the analysis of the electron dynamics in the different simulations, we would like to point out some distinct features of the bubbles and the structure of the fields inside them. In the composite vortex beams case (simulation A), the laser intensity lobes are bean-shaped in the transverse plane unlike in Gaussian driven wakefields. The same transverse shape is observed in the ion cavity behind it. Importantly, in the rotating case, there is an additional magnetic field component, $B_1$, shown in Figure \ref{fig:flds_rot_wake_bubbles}. 

\begin{figure}[ht]
\centering
\includegraphics[width=\linewidth]{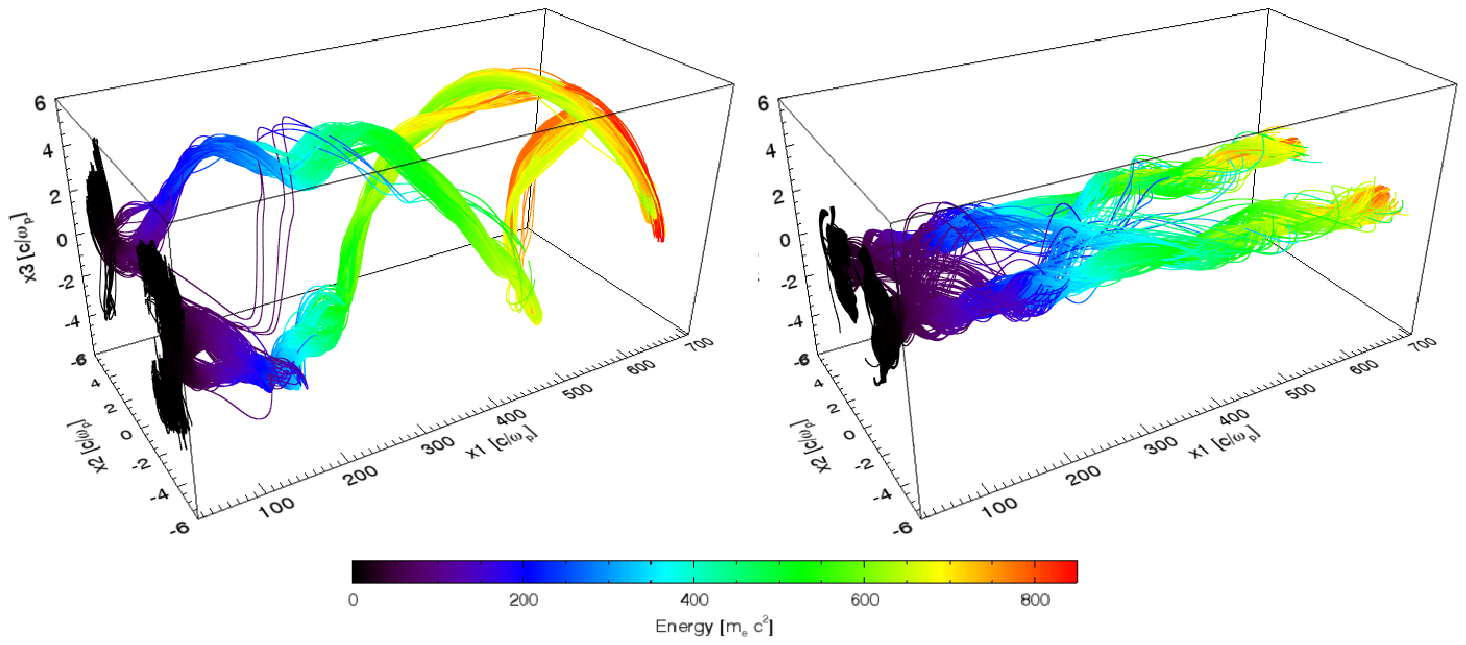} 
\caption{Sample of trajectories from the rotating wakefield (left panel, 1446 particles) and non-rotating (right panel, 992 particles) wakefields scenarios, coloured according to the energy.}
\label{fig:tracks_rot_non_rot}
\end{figure}

In all the simulations, electrons are injected through ionisation injection and some become trapped. Samples of these electrons were taken from each simulation. In the rotating wakefield scenario (A), trapped electrons were observed from all the Nitrogen ionisation levels and a negligible number from Helium atoms. In the non-rotating case (B), however, only electrons from the $6^{th}$ and $7^{th}$ levels were trapped in non-negligible number. The sample electron trajectories are depicted in Figure \ref{fig:tracks_rot_non_rot}. In the Gaussian driver case (C), only electrons from $6^{th}$ and $7^{th}$ levels were trapped at this time, though later electrons from other levels and from the Helium atoms were also injected. This means the sample is taken from the electrons that will first reach the highest energies since ionisation injection occurs over an extended period of time. 

The most striking difference in the trajectories between scenarios A and B is the helical motion observed in the rotating wakefields (left panel in Figure \ref{fig:tracks_rot_non_rot}). As expected, the wakefield bubble rotation drags the trapped electrons with it, leading to a low-frequency helix-type trajectory. The radius of this helix can be estimated from the distance from the centroids of the laser lobes to the axis of propagation \cite{Jorge_arXiv_2018}:
\begin{equation}
R_h \simeq W_0 \sqrt{|\ell_0|/2},    
\label{eq_Rh}
\end{equation}
where $\ell_0 = (\ell_1 + \ell_2) / 2$. Superimposed on this helix, betatron oscillations are clearly visible in the early stages of the trajectory but become smoothed out later on,  and the azimuthal motion dominates the dynamics of the electrons. Since the electron energy is also increasing with time, the latter part of the trajectory will contribute the most to the radiation emission, as usual in laser-wakefield accelerators. This suggests that the radiation emission properties will mostly be determined by the helical motion. 

\begin{figure}[ht]
\centering
\includegraphics[width=\linewidth]{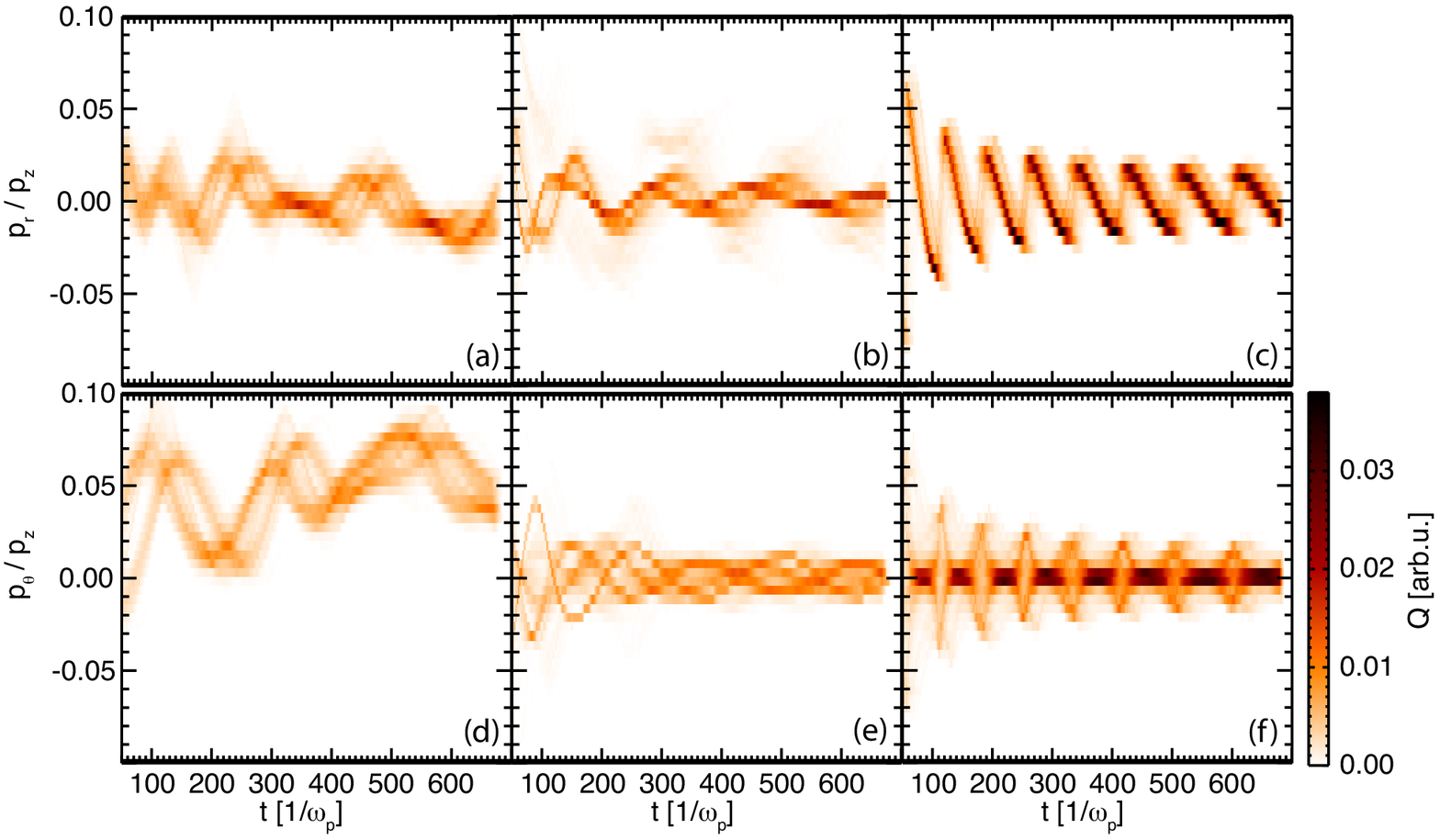}
\caption{Temporal evolution of the momentum radial and azimuthal components of the trapped electron sample from: LWFA driven by composite vortex beams with $\ell_1 = 3$, $\ell_2 = 5$ (a,d) and $\ell_1 = 1$, $\ell_2 = -1$ (b,e) and by a Gaussian laser beam (c,f).}
\label{fig:pr_pth_all_cases}
\end{figure}

Indeed, the rotation imposed by the laser has important consequences on the transverse dynamics. As observed in Figure \ref{fig:pr_pth_all_cases}, while the amplitude of the radial momentum is of similar amplitude in all three scenarios, the azimuthal momentum in the rotating wakefield is up to four times higher than in the other two scenarios. Since the acceleration in the direction perpendicular to the momentum contributes the most to the radiated power~\cite{Jackson}, an enhancement of the radiated energy, over the non-rotating cases, should be expected.
\subsection*{Radiation emission}

To investigate the impact of the rotation on the radiation emission properties, the trajectory samples were post-processed with the code jRad \cite{jrad}.

\begin{figure}[ht]
\centering
\includegraphics[width=\linewidth]{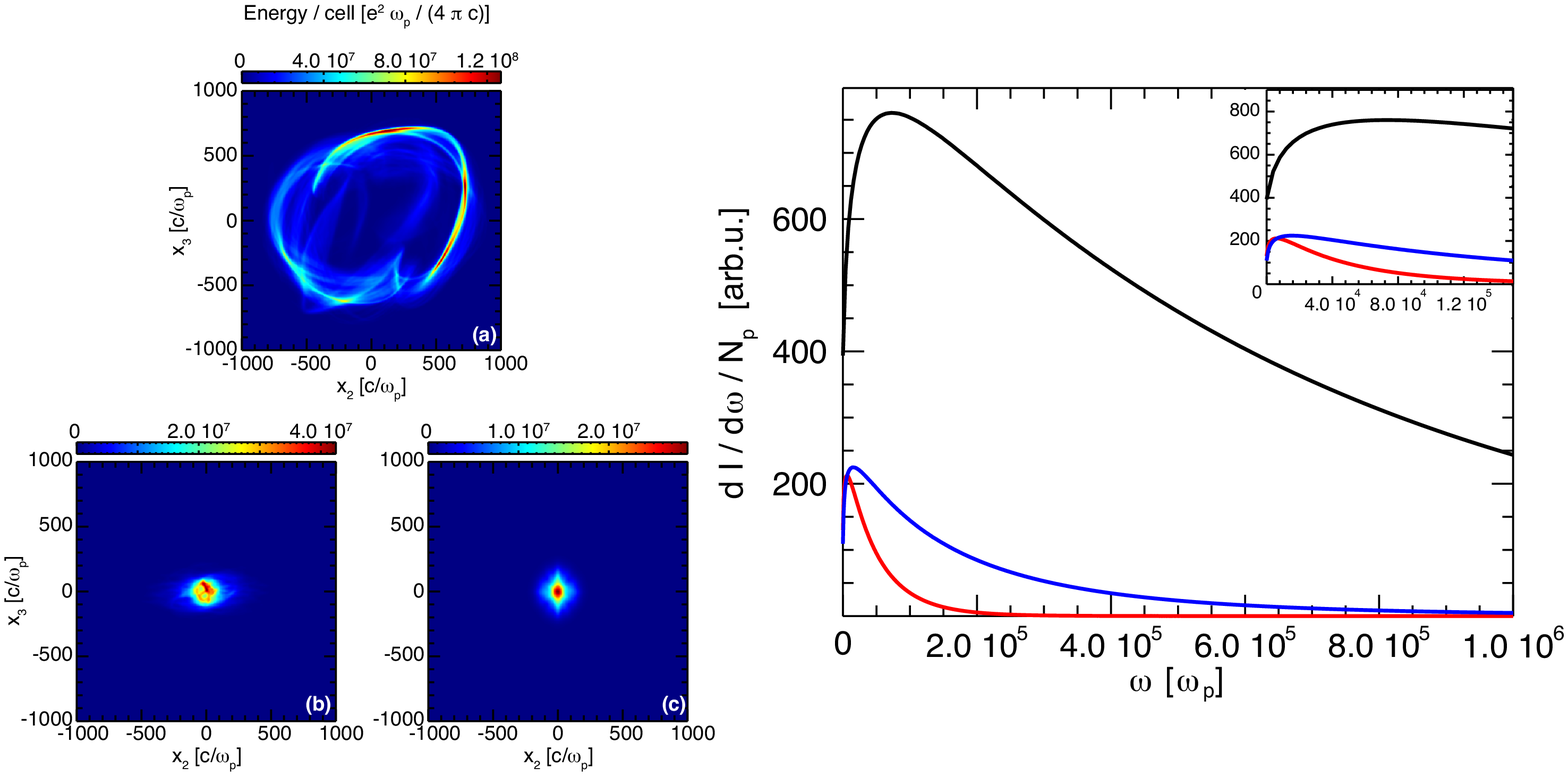}
\caption{(Left panels) Energy radiated by the trapped electron sample on a detector placed a distance $d = 10^4 \, c/\omega_p$ away from the origin, perpendicularly to the propagation axis. Panels (a,b,c) refer to the rotating wakefields, non-rotating wakefields and Gaussian laser driver scenarios respectively. (Right panel) Spectrum of the radiation emitted by the trapped electron sample integrated over the detector area for the rotating wakefields (black), non-rotating wakefields (blue) and Gaussian (red) laser driver scenarios respectively. Each curve is divided by the number of electrons in the respective sample. The inset shows a zoom in the lower frequency part of the spectrum.}
\label{fig:ene_incoh_dIdw_per_e_3cases}
\end{figure}

For all three scenarios explored in the simulations, the radiated energy emitted by each trapped electron sample (and calculated incoherently) was determined (see Figure \ref{fig:ene_incoh_dIdw_per_e_3cases}). As expected, the emission from the electrons undergoing helical motion observed in the rotating wakefield scenario traces an approximately annular profile in the detector. The spatial profile of the radiated energy in the other two scenarios is similar to typical LWFA energy patterns, with most of the energy at the centre. 

The radiated energy captured in the detector (in Figure \ref{fig:ene_incoh_dIdw_per_e_3cases}) was integrated over the area covered by the detector to obtain the total radiated energy. The electron samples analysed are a small fraction of the total electron bunches. For this reason, the most interesting value in this study is the energy radiated per electron, which was determined to be $\simeq 27 \,\mathrm{keV}$, $\simeq 2 \,\mathrm{keV}$ and $\simeq 0.5 \,\mathrm{keV}$ for the rotating wakefields, non-rotating wakefields and Gaussian laser driver scenarios respectively. This represents a factor of approximately thirteen times more radiated energy in the rotating wakefield case compared to the non-rotating wakefield over the same propagation distance.

The radiated energy spectrum, integrated over the detector area, is shown in Figure \ref{fig:ene_incoh_dIdw_per_e_3cases}. It is observed that in the rotating wakefield scenario, the spectrum has a peak at a frequency approximately four times higher than in the non-rotating case. 

The dynamics of electrons undergoing betatron (or helical) oscillations is analogous to those experienced in wigglers, i.e.~magnetic field devices composed of arrays of magnets of alternating polarity. Both in magnetic wigglers and plasma wigglers, information about the main properties of the radiation can be obtained from quantities such as the wiggler strength parameter $K$ (the ratio between the maximum angle of deflection of the trajectory to the angular aperture of radiation),  the fundamental frequency $\omega_f$, and the critical frequency $\omega_{cr}$\cite{Esarey_PRE_2002,Corde_RevModPhys_2013}:
\begin{subequations}
\begin{eqnarray}
K &=& \sqrt{\gamma / 2} r_w \\
\hbar \omega_f &=& (2\gamma^2 h c / \lambda_w) / (1 + K^2/2) \\
\hbar \omega_{cr} &=&  (3/2) K \gamma^2 h c / \lambda_w,
\end{eqnarray}
\label{eqs_wigg_params}
\end{subequations}
where $r_w$ is the amplitude of the oscillations (for $K \gg 1$). For strongly relativistic electrons and a number of oscillations $N_{\beta} \gg 1$, the radiated spectrum has been shown to asymptotically tend to the synchrotron spectrum \cite{Esarey_PRE_2002} and the average radiated power (averaged over one period) is $\bar{P}_{\gamma,d} = [e^2 \omega_p / (m_e c^3)] \gamma^2 K^2 {\omega_w}^2 / (12 \pi)$ \cite{Corde_RevModPhys_2013}. For $K\gg 1$, the radiation spectrum is composed of many harmonics and extends up to the critical frequency $\omega_{cr}\simeq 3 K \gamma^2 \omega_\beta$, where $\omega_{\beta} = \omega_p / \sqrt{2\gamma}$ is the betatron oscillation frequency, and decays exponentially afterwards.

In laser-wakefield accelerators, wiggler emission also occurs, but the presence of an accelerating electric field in the direction of the laser propagation makes these parameters time-dependent \cite{Thomas_PoP_2010,Corde_RevModPhys_2013}. Nevertheless, they provide useful information on the local properties of the radiation, and experimental results have shown that the spectrum is still approximately synchrotron-like \cite{Fourmaux_NJP_2011}.

In the scenarios studied in this work, if one interprets the spectrum in terms of the wiggler  analogy, the local critical frequency is (apart from the electron density) just a function of the particle energy and amplitude of oscillation ($r_w$), $\omega_{cr} = (3 / 4) \gamma^2 r_w \omega_p^2 / c$. The average Lorentz factor of the trapped electrons at the end of the propagation is about $\gamma \simeq 700$ for both LG pulse scenarios and $\gamma \simeq 450$ for the Gaussian driver scenario. However, the amplitude of oscillation differs substantially. While in the non-rotating and the Gaussian driver case it is given by the local betatron oscillation amplitude $r_{\beta}$, in the rotating scenario it is approximately equal to the distance from the centre of each bubble to the propagation axis $R_{h}$. Visual inspection of the projection of the trajectories in the plane transverse to the laser propagation yields $R_{h} \sim 4 \, c/\omega_p$, which is in good agreement with the theoretical value of $R_h = 4.24  \, c/\omega_p$ (from Eq. \ref{eq_Rh}), and $r_{\beta} \sim 1 \, c/\omega_p$. This is consistent with the observed differences in the spectrum between the rotating and non-rotating cases.

Considering that the radiated spectrum critical frequency has a $\gamma^2$ dependence, one  expects the Gaussian-driver LWFA spectrum to extend up to a factor $\sim (700/450)^2 \simeq 2.5$ less than for the composite vortex beam case. From the observation of the radiated spectrum in the right panel of Figure \ref{fig:ene_incoh_dIdw_per_e_3cases}, the ratio between the extension of the spectra for the Gaussian driver is approximately a factor of three lower than in the non-rotating composite vortex wakefields. This suggests that the non-rotating LG case does not significantly enhance the radiation energy compared to the Gaussian case.

\subsection*{Tuning the radiation properties}

The second advantage of our proposed setup is that it allows greater control of the radiated energy spectrum compared to typical LWFA betatron sources. 
A qualitative comparison between the main features of betatron radiation from the rotating wakefields studied in this work and betatron radiation from typical laser-wakefields can be performed by inspecting the ratio between the expressions for the main parameters of the radiation in the two cases. To facilitate comparisons, in the following we present these ratios in normalised units, with frequencies normalised to $\omega_p$, lengths to $c/\omega_p$ and energy to $m_e c^2$. Normalised quantities will appear with tildes to distinguish them from the previous occurrences in the text.

In typical LWFA betatron radiation the wavelength of the oscillation is $\lambda_{\beta} = 2\pi c / \omega_{\beta}$, where $\omega_{\beta} = \omega_p / \sqrt{2\gamma}$ is the betatron oscillation frequency \cite{Esarey_PRE_2002} and the amplitude of oscillation scales with the electron energy as $r_{\beta} = {r_{\beta}}_0 \gamma_0^{1/4} / \gamma(t)^{1/4}$ ~\cite{Glinec_EurPhysLett_2008}. Furthermore, $\gamma > \gamma_{\phi}$ for electrons to be trapped, so the initial Lorentz factor of the electrons inside the bubble can be approximated by $\gamma_0 \sim \gamma_{\phi}$, which for very underdense plasmas is $\gamma_{\phi} \simeq \tilde{\omega}_0 / \sqrt{3}$. Substituting these in the wiggler parameters yields:
\begin{subequations}
\begin{eqnarray}
\tilde{\omega}_f &\simeq& \frac{4(3^{1/4})\sqrt{2}}{ {\tilde{r}_{\beta_0}}^2 \sqrt{\tilde{\omega}_0} }\gamma, \\
\tilde{\omega}_{cr} &=&  \frac{3^{7/4}}{4} \tilde{r}_{\beta_0} \tilde{\omega}_0^{1/4} \gamma^{7/4}, \\
K &=& \frac{1}{3^{1/8}\sqrt{2}} \tilde{r}_{\beta_0} \tilde{\omega}_0^{1/4} \gamma^{1/4},
\end{eqnarray}
\label{wigg_params_nonrot_lwfa}
\end{subequations}
where the approximation $\gamma \,\tilde{r}^2 \gg 1$ was used.

In the rotating wakefield scenario, as the dynamics is dominated by the helical motion, the wiggler amplitude of oscillation is given by the radius of the helix $R_h$ and the wavelength of the oscillation is $\lambda = 2 \pi c / \Omega_h$. Using the expressions for $\tilde{R}_h$ and $\tilde{\Omega}_h$ from the previous section, this yields:
\begin{subequations}
\begin{eqnarray}
\tilde{\omega}_f &\simeq& \frac{8 \gamma |\tilde{\Omega}_h|}{R_h^2}  = \frac{32 \gamma}{\tilde{\omega}_0 \tilde{W}_0^4 |\ell_0|}, \\
\tilde{\omega}_{cr} &=&  \frac{3 \sqrt{2}}{4} \tilde{R}_h |\tilde{\Omega}_h| \gamma^{5/2} = \frac{3}{2}\frac{\sqrt{\ell_0} \gamma^{5/2}}{\tilde{\omega}_0 \tilde{W}_0},  \\
K &=& \sqrt{\gamma / 2} \tilde{R}_h = \tilde{W}_0 \sqrt{\gamma |\ell_0|} / 2.
\end{eqnarray}
\label{wigg_params_rot_lwfa}
\end{subequations}

The ratio between the parameters of the two scenarios, (\ref{wigg_params_nonrot_lwfa}) and (\ref{wigg_params_rot_lwfa}), is then:
\begin{subequations}\label{eq:ratios_rot_nonrot}
\begin{eqnarray}
\frac{{\tilde{\omega}_f}^{rot.}}{{\tilde{\omega}_f}^{non-rot.}} &\simeq& 4.3 \frac{{\tilde{r}_{\beta_0}}^2}{ \tilde{\omega}_0^{1/2}} \frac{1}{\tilde{W}_0^4 |\ell_0|}, \label{eq:ratio_wf}\\
\frac{{\tilde{\omega}_{cr}}^{rot.}}{{\tilde{\omega}_{cr}}^{non-rot.}} &=& \frac{\tilde{R}_h}{\tilde{r}_{\beta}} \frac{|\tilde{\Omega}_h|}{\tilde{\omega}_{\beta}} = 2.3\frac{\sqrt{|\ell_0|}}{\tilde{r}_{\beta_0} \tilde{W}_0} \frac{\gamma^{3/4}}{\tilde{\omega}_0^{5/4}}, \label{eq:ratio_wcr} \\
\frac{K^{rot.}}{K^{non-rot.}} &=& \frac{\tilde{R}_h}{\tilde{r}_{\beta}} = 0.8 \frac{\tilde{W}_0 \sqrt{|\ell_0|}}{\tilde{r}_{\beta_0} } \left(\frac{\gamma}{\tilde{\omega}_0}\right)^{1/4}. \label{eq:ratio_K}
\end{eqnarray}
\end{subequations}

It is interesting to note that, for the simulation parameters employed here, the fundamental frequency and the wiggler parameter for the rotating and non-rotating cases are quite different. If we approximate the initial betatron radial position in the non-rotating case to be a fraction of the laser spot size $r_{\beta_0} \simeq \alpha W_0$ (and assume that the spot size is of the same order in all the cases), we can estimate these two parameters as a function of $\gamma$ and $\alpha$ only. Taking the approximate average final value of $\gamma_f \simeq 700$, fundamental frequencies of $\omega_f^{rot.} \simeq 2.3 \, \omega_p$ and $\omega_f^{non-rot.} \simeq (60 / \alpha^2) \, \omega_p$ are obtained, respectively, where $\alpha < 1$. Using the same $\gamma$, the wiggler strength parameter is $K^{rot}\simeq 80$ and $K^{non-rot.}\simeq 30 \alpha \ll K^{rot}$, respectively. This means the rotating wakefield scenario radiation emission occurs in a distinct regime, where the fundamental frequency is much lower than in a typical wakefield with similar laser parameters, but the wiggler strength parameter is much higher. If the ratio of the critical frequencies is expressed in terms of the wiggler parameters (using Eqs. (\ref{eq:ratio_wcr}) and (\ref{eq:ratio_K})):
\begin{equation}
\frac{{\tilde{\omega}_{cr}}^{rot.}}{{\tilde{\omega}_{cr}}^{non-rot.}} = \frac{K^{rot.}}{K^{non-rot.}} \frac{|\tilde{\Omega}_h|}{\tilde{\omega}_{\beta}},
\label{eq_cr_K}\end{equation}
it also helps to understand why the critical frequency is still significantly higher for the rotating laser scenario despite the much lower oscillation amplitude.

Finally, the ratio between the average instantaneous radiated power $\bar{P}_{\gamma}$ in both cases is given by:
\begin{equation}
\frac{{\bar{P}_{\gamma}}^{rot.}}{{\bar{P}_{\gamma}}^{non-rot.}} \simeq 5.3 \frac{|\ell_0|}{\tilde{W}_0^2 \tilde{r}_{\beta_0}^2 \tilde{\omega}_0^{5/2}} \gamma^{3/2}.
\label{eq_power}\end{equation}

From these expressions, one can determine for which parameters of the laser the betatron emission in rotating wakefields is more advantageous than in the non-rotating case, and how to tune the parameters to achieve the desired radiation properties.  It should be noted, however, that in these expressions it is assumed that the electron energy scales with propagation distance (or time) in the same manner in both the rotating and non-rotating scenario. Dephasing was observed to occur sooner in the rotating case compared to the non-rotating one in the simulations of the previous section. While this means that electrons injected in these cases could reach higher energies if the propagation of the laser drivers were extended, it does not follow that radiation emission would surpass that of the rotating laser scenario since the oscillation amplitude scales with  $\gamma^{-1/4}$.

An additional approximation in this analysis is that the parameters associated with the laser-driver rotation, hence the wakefield rotation, are considered constant. As depletion of the laser sets in, the wakefield group velocity and angular velocity will change.

In laser-wakefields, the plasma is underdense and the initial radial position of trapped electrons is less than the bubble radius, which is approximately equal to the laser spot size under matched conditions. Furthermore, the laser spot size is typically larger than $c/\omega_p$. This suggests that for the same plasma density and similar laser spot sizes in the rotating and non-rotating cases, if $\gamma \gg \tilde{\omega}_0$ (which is valid apart from the earlier stages of the trajectories), the wiggler strength parameter should be higher in the rotating case according to equation (\ref{eq:ratio_K}). Under the same considerations, the fundamental frequency of the radiation emitted by the trapped electrons will be much lower in rotating case.  This indicates that the distinct regimes between rotating and non-rotating scenarios for wakefields driven by laser drivers with similar parameters, which were seen in the previous section, should be observed in general.

\section*{Conclusion}

In this work, the possibility of exploiting composite vortex beams to drive rotating wakefields and enhance the radiation emission from the accelerated electrons was explored. The composite vortex laser driver can lead to rotating wakefields in the plasma. The electrons trapped in these bubbles are thus forced to follow this imposed helical motion.
This motion can lead to enhanced radiation emission compared to betatron radiation from typical laser-wakefield accelerators. 

Our simulation results show over ten times higher energy emitted per electron in the case of rotating wakefields compared to non-rotating wakefields driven by either non-rotating composite or Gaussian laser beam drivers. These simulations were performed with the same pulse energy for comparison purposes. We have also demonstrated that, even though the fundamental frequency of radiation in the rotating case is generally much lower than the betatron frequency, the wiggler strength parameter will be higher. For the parameters chosen here, the critical frequency was also higher than in the comparable non-rotating scenarios. This shows that, for the same laser energy and comparable laser parameters, choosing a rotating wakefield setup driven by a composite vortex beam can provide a much more efficient and higher-frequency radiation source, albeit with higher divergence.

In typical laser-wakefield scenarios the properties of the betatron radiation are determined by the plasma properties, after matching conditions are imposed. However, the characteristics of radiation emitted by the electrons accelerated in rotating wakefields can be tuned through adjusting the spot size and orbital angular momentum indexes. As these are not fixed by the matching conditions, this provides a novel capability -- tuning of these radiation sources.

\section*{Methods}

\subsection*{Particle-in-cell simulations}

The three-dimensional particle-in-cell (PIC) simulations in this work were performed within the OSIRIS framework \cite{osiris1,osiris2}. In the first two simulations (A and B), the laser driver is composed of two LG pulses, each with normalised peak vector potential $a_0 = 1.5$ (giving a peak $a_0 \sim 3$ for the vortex structure), frequency ${\omega_0}_{(1,2)} = 30, 30.01 \, \omega_p$, pulse length $\tau_0 = 2 \, \omega_p^{-1}$ and spot size ${W_0}_{(1,2)} = 3, \, 4.1 \, c/\omega_p$. The LG radial index is $p=0$ in all cases and the azimuthal indexes are $\ell_1 = 3$ and $\ell_2 = 5$ in simulation A, and $\ell_1 = 1$ and $\ell_2 = -1$ in simulation B. In the third simulation, which is referred to as simulation C in the text, the driver is a standard Gaussian laser pulse with $a_0 = 2.5$, frequency $\omega_0 = 30 \, \omega_p$, pulse length $\tau_0 = 2 \, \omega_p^{-1}$ and spot size $W_0 = 5.7 \, c/\omega_p$. 

The simulation box is $12 \times 20 \times 20 \, (c/\omega_p)^3$, decomposed into a grid with $1800\times 200\times 200$ cells, and the time resolution is $\Delta t = 3.8 \times 10^{-3} \omega_p^{-1}$. Ionisation is modelled with the ADK rates \cite{ADK_JETP_1986}, two macro-particles per ion are used for the electrons from the Helium atoms and ten macro-particles per ion for the Nitrogen electrons.
The sample electrons for radiation post-processing were chosen randomly (from the electrons above a given energy threshold between $\gamma = 30$ and $60$, such that they are expected to be trapped) early in the simulation, at $t \simeq 60\, \omega_p^{-1}$.

In each case, the laser driver propagates in a plasma channel with a transverse parabolic density profile with a channel radius of $r_0 = k_p W_0$. The longitudinal profile begins with a $12 \, c/\omega_p$ length linear ramp starting at $x_1 = 13 \, c/\omega_p$ and then stays flat until the end of the propagation distance, at $x_1 \simeq 712 \, c/\omega_p$. The plasma is composed of a mixture of Hydrogen ($n_0 = 1.94 \times 10^{18} \, \mathrm{cm}^{-3}$) with a small percentage of Nitrogen ($n_N = 10^{-5} n_0$) to achieve ionisation injection \cite{Pak_PRL_2010,McGuffey_PRL_2010}. This avoids the need for external injection, which would require fine control of the initial beam parameters.

\subsection*{Radiation calculation}

The radiation emission analysis in this work was performed using the jRad post-processing radiation code \cite{jrad}. The radiated energy and spectrum are determined (incoherently) using the standard formulas from classical electrodynamics \cite{Jackson}:
\begin{eqnarray}
E &=& \frac{e^2}{4 \pi c} \sum_p \int_{-\infty}^{+\infty} \frac{| \mathbf{n}\times [(\mathbf{n}-\boldsymbol{\beta})\times \dot{\boldsymbol{\beta}}] |^2}{(1 - \mathbf{n} \cdot \boldsymbol{\beta})^5 R^2} S_{pixel} ~dt' \\
\frac{d^2 I}{d\omega d\Omega} &=& \frac{e^2 \omega_p^2}{4 \pi^2 c^3} \sum_p \left| \int_{-\infty}^{+\infty} \frac{\bold{n}\times \{ (\bold{n}-\boldsymbol{\beta})\times \dot{\boldsymbol{\beta}} \}}{(1-\bold{n}\cdot\boldsymbol{\beta})^2} \exp [i \omega (t'+R(t')/c)] dt' \right|^2
\end{eqnarray}
where the summation in $p$ is in the particles of the sample, $S_{pixel}$ is the detector cell area, $\boldsymbol{\beta}$ is the velocity of the electron normalised to the speed of light $c$ and $\dot{\boldsymbol{\beta}}$ its time derivative, $R$ is the distance from the particle position to the center of the detector cell, $\mathbf{n}$ is the unit vector pointing from the particle position to the center of the detector cell. The evolution of the electron sample in phasespace is obtained by recording the trajectories of these electrons during the PIC simulation. Samples of 992 1446 and 2000 macroparticles were recorded in simulations A, B and C respectively. A time-resolution of $4 \Delta t$ and detector grid with $400\times 400$ cells were employed in the energy calculation. In the spectrum detector a resolution of $104\times 104$ was used in space and $1000$ points in the frequency axis, and a time resolution of $4 \Delta t$ was employed. The spatial region used for the spectrum calculation was $10^3\times 10^3 ~(c/\omega_p)^2$ for simulation A and $300\times 300 ~(c/\omega_p)^2$ for simulations B and C, while the maximum frequency considered was $4\times 10^6 ~\omega_p$ for simulation A and $10^6 ~\omega_p$ for simulations B and C. Both the energy and spectrum detectors were positioned  perpendicularly to the laser propagation axis and placed at $x_1 = 10^4 \, c/\omega_p$.

\subsection*{Momentum in cylindrical coordinates}

Figure \ref{fig:pr_pth_all_cases} was obtained by first taking the electron trajectories sample in each simulation in cartesian coordinates and determining the radial and azimuthal momentum divided by the momentum in the laser propagation direction. The radial coordinate was calculated relative to the laser propagation axis. The macroparticle charge was then deposited in a grid where the time axis has the same resolution as the trajectories ($4 \Delta t$) and the momentum bin size is $\Delta p = 0.005 ~m_e c$.

\subsection*{Data availability} The data that support the findings of this study are available from the corresponding author upon request.



\section*{Acknowledgements}

The authors would like to acknowledge fruitful discussions with R.~Fonseca, I.~Abel, L.O.~Silva and L.~Yi. This work was supported by the European Research Council (ERC-2014-CoG grant 647121) and the Knut and Alice Wallenberg Foundation. J.~V. acknowledges the support of FCT (Portugal) Grant No. SFRH/IF/01635/2015. The authors acknowledge the grant of computing time by the Leibnitz Research Center on SuperMUC, where the simulations were performed.

\section*{Author contributions statement}

J.L.M designed and conducted the simulations. J.L.M analysed the data in collaboration with J.V., J.F. and T.F.. J.L.M and T.F. wrote the manuscript with contributions from J.V. and J.F..

\section*{Additional information}

\textbf{Competing interests} The authors declare no competing interests.


\begin{thebibliography}{10}
\expandafter\ifx\csname url\endcsname\relax
  \def\url#1{\texttt{#1}}\fi
\expandafter\ifx\csname urlprefix\endcsname\relax\def\urlprefix{URL }\fi
\expandafter\ifx\csname doiprefix\endcsname\relax\def\doiprefix{DOI }\fi
\providecommand{\bibinfo}[2]{#2}
\providecommand{\eprint}[2][]{\url{#2}}

\bibitem{Cole_SciRep_2004}
\bibinfo{author}{Cole, J.} \emph{et~al.}
\newblock \bibinfo{journal}{\bibinfo{title}{Laser-wakefield accelerators as
  hard x-ray sources for 3d medical imaging of human bone}}.
\newblock {\emph{\JournalTitle{Sci Rep.}}} \textbf{\bibinfo{volume}{5}},
  \bibinfo{pages}{13244} (\bibinfo{year}{2015}).

\bibitem{Albert_PPCF_2016}
\bibinfo{author}{Albert, F.} \& \bibinfo{author}{Thomas, A.}
\newblock \bibinfo{journal}{\bibinfo{title}{Applications of laser wakefield
  accelerator-based light sources}}.
\newblock {\emph{\JournalTitle{Plasma Phys. Contr. F.}}}
  \textbf{\bibinfo{volume}{58}}, \bibinfo{pages}{103001}
  (\bibinfo{year}{2016}).

\bibitem{Albert_ProcSpie_2017}
\emph{\bibinfo{title}{X-ray absorption spectroscopy of warm dense matter with
  betatron x-ray radiation (Conference Presentation)}}, vol.
  \bibinfo{volume}{10243}.

\bibitem{TajimaDawson_PRL_1979}
\bibinfo{author}{Tajima, T.} \& \bibinfo{author}{Dawson, J.~M.}
\newblock \bibinfo{journal}{\bibinfo{title}{Laser electron accelerator}}.
\newblock {\emph{\JournalTitle{Phys. Rev. Lett.}}}
  \textbf{\bibinfo{volume}{43}}, \bibinfo{pages}{267} (\bibinfo{year}{1979}).

\bibitem{Geddes_Nature_2004}
\bibinfo{author}{Geddes, C.} \emph{et~al.}
\newblock \bibinfo{journal}{\bibinfo{title}{High-quality electron beams from a
  laser wakefield accelerator using plasma-channel guiding}}.
\newblock {\emph{\JournalTitle{Nature (London)}}}
  \textbf{\bibinfo{volume}{431}}, \bibinfo{pages}{538} (\bibinfo{year}{2004}).

\bibitem{Mangles_Nature_2004}
\bibinfo{author}{Mangles, S.} \emph{et~al.}
\newblock \bibinfo{journal}{\bibinfo{title}{Monoenergetic beams of relativistic
  electrons from intense laser-plasma interactions}}.
\newblock {\emph{\JournalTitle{Nature (London)}}}
  \textbf{\bibinfo{volume}{431}}, \bibinfo{pages}{535-538}
  (\bibinfo{year}{2004}).

\bibitem{Faure_Nature_2004}
\bibinfo{author}{Faure, J.} \emph{et~al.}
\newblock \bibinfo{journal}{\bibinfo{title}{A laser-plasma accelerator
  producing monoenergetic electron beams}}.
\newblock {\emph{\JournalTitle{Nature (London)}}}
  \textbf{\bibinfo{volume}{431}}, \bibinfo{pages}{541-544}
  (\bibinfo{year}{2004}).

\bibitem{Leemans_NatPhys_2006}
\bibinfo{author}{Leemans, W.~P.} \emph{et~al.}
\newblock \bibinfo{journal}{\bibinfo{title}{Gev electron beams from a
  centimetre-scale accelerator}}.
\newblock {\emph{\JournalTitle{Nat. Phys.}}} \textbf{\bibinfo{volume}{2}},
  \bibinfo{pages}{696} (\bibinfo{year}{2006}).

\bibitem{Clayton_PRL_2010}
\bibinfo{author}{Clayton, C.~E.} \emph{et~al.}
\newblock \bibinfo{journal}{\bibinfo{title}{Self-guided laser wakefield
  acceleration beyond 1 gev using ionization-induced injection}}.
\newblock {\emph{\JournalTitle{Phys. Rev. Lett.}}}
  \textbf{\bibinfo{volume}{105}}, \bibinfo{pages}{105003}
  (\bibinfo{year}{2010}).

\bibitem{Leemans_PRL_2014}
\bibinfo{author}{Leemans, W.~P.} \emph{et~al.}
\newblock \bibinfo{journal}{\bibinfo{title}{Multi-gev electron beams from
  capillary-discharge-guided subpetawatt laser pulses in the self-trapping
  regime}}.
\newblock {\emph{\JournalTitle{Phys. Rev. Lett.}}}
  \textbf{\bibinfo{volume}{113}}, \bibinfo{pages}{245002}
  (\bibinfo{year}{2014}).

\bibitem{Hyung_Taek_SciRep_2017}
\bibinfo{author}{Kim, H.~T.} \emph{et~al.}
\newblock \bibinfo{journal}{\bibinfo{title}{Stable multi-gev electron
  accelerator driven by waveform-controlled pw laser pulses}}.
\newblock {\emph{\JournalTitle{Sci. Rep.}}} \textbf{\bibinfo{volume}{7}},
  \bibinfo{pages}{10203} (\bibinfo{year}{2017}).

\bibitem{Whittum_POF_1992}
\bibinfo{author}{Whittum, D.~H.}
\newblock \bibinfo{journal}{\bibinfo{title}{Electromagnetic instability of the
  ion-focused regime}}.
\newblock {\emph{\JournalTitle{Phys. Fluids. B - Plasma}}}
  \textbf{\bibinfo{volume}{4}}, \bibinfo{pages}{730--739}
  (\bibinfo{year}{1992}).

\bibitem{Esarey_PRE_2002}
\bibinfo{author}{Esarey, E.}, \bibinfo{author}{Shadwick, B.},
  \bibinfo{author}{Catravas, P.} \& \bibinfo{author}{Leemans, W.}
\newblock \bibinfo{journal}{\bibinfo{title}{Synchrotron radiation from electron
  beams in plasma-focusing channels}}.
\newblock {\emph{\JournalTitle{Phys. Rev. E}}} \textbf{\bibinfo{volume}{65}},
  \bibinfo{pages}{056505} (\bibinfo{year}{2002}).

\bibitem{Rousse_PRL_2004}
\bibinfo{author}{Rousse, A.} \emph{et~al.}
\newblock \bibinfo{journal}{\bibinfo{title}{Production of a kev x-ray beam from
  synchrotron radiation in relativistic laser-plasma interaction}}.
\newblock {\emph{\JournalTitle{Phys. Rev. Lett.}}}
  \textbf{\bibinfo{volume}{93}}, \bibinfo{pages}{135005}
  (\bibinfo{year}{2004}).

\bibitem{TaPhuoc_PoP_2005}
\bibinfo{author}{Ta~Phuoc, K.} \emph{et~al.}
\newblock \bibinfo{journal}{\bibinfo{title}{Laser based synchrotron
  radiation}}.
\newblock {\emph{\JournalTitle{Phys. Plasmas}}} \textbf{\bibinfo{volume}{12}},
  \bibinfo{pages}{023101} (\bibinfo{year}{2005}).

\bibitem{Kneip_2010}
\bibinfo{author}{Kneip, S.} \emph{et~al.}
\newblock \bibinfo{journal}{\bibinfo{title}{Bright spatially coherent
  synchrotron x-rays from a table-top source}}.
\newblock {\emph{\JournalTitle{Nature (London)}}}
  \textbf{\bibinfo{volume}{431}}, \bibinfo{pages}{541-544}
  (\bibinfo{year}{2004}).

\bibitem{Corde_RevModPhys_2013}
\bibinfo{author}{Corde, S.} \emph{et~al.}
\newblock \bibinfo{journal}{\bibinfo{title}{Femtosecond x rays from
  laser-plasma accelerators}}.
\newblock {\emph{\JournalTitle{Rev. Mod. Phys.}}}
  \textbf{\bibinfo{volume}{85}}, \bibinfo{pages}{1--48} (\bibinfo{year}{2013}).

\bibitem{Lu_PRSTAB_2016}
\bibinfo{author}{Lu, W.} \emph{et~al.}
\newblock \bibinfo{journal}{\bibinfo{title}{Generating multi-gev electron
  bunches using single stage laser wakefield acceleration in a 3d nonlinear
  regime}}.
\newblock {\emph{\JournalTitle{Phys. Rev. Spec. Top.-Ac.}}}
  \textbf{\bibinfo{volume}{10}}, \bibinfo{pages}{061301}
  (\bibinfo{year}{2007}).

\bibitem{Corde_PRL_2011}
\bibinfo{author}{Corde, S.} \emph{et~al.}
\newblock \bibinfo{journal}{\bibinfo{title}{Controlled betatron x-ray radiation
  from tunable optically injected electrons}}.
\newblock {\emph{\JournalTitle{Phys. Rev. Lett.}}}
  \textbf{\bibinfo{volume}{107}}, \bibinfo{pages}{255003}
  (\bibinfo{year}{2011}).

\bibitem{Vieira_PPCF_2012}
\bibinfo{author}{Vieira, J.} \emph{et~al.}
\newblock \bibinfo{journal}{\bibinfo{title}{Magnetically assisted
  self-injection and radiation generation for plasma-based acceleration}}.
\newblock {\emph{\JournalTitle{Plasma Phys. Contr. F.}}}
  \textbf{\bibinfo{volume}{54}}, \bibinfo{pages}{124044}
  (\bibinfo{year}{2012}).

\bibitem{Rykovanov_PRL_2015}
\bibinfo{author}{Rykovanov, S.~G.}, \bibinfo{author}{Schroeder, C.~B.},
  \bibinfo{author}{Esarey, E.}, \bibinfo{author}{Geddes, C. G.~R.} \&
  \bibinfo{author}{Leemans, W.~P.}
\newblock \bibinfo{journal}{\bibinfo{title}{Plasma undulator based on laser
  excitation of wakefields in a plasma channel}}.
\newblock {\emph{\JournalTitle{Phys. Rev. Lett.}}}
  \textbf{\bibinfo{volume}{114}}, \bibinfo{pages}{145003}
  (\bibinfo{year}{2015}).

\bibitem{Luo_SciRep_2016}
\bibinfo{author}{Luo, J.} \emph{et~al.}
\newblock \bibinfo{journal}{\bibinfo{title}{A compact tunable polarized x-ray
  source based on laser-plasma helical undulators}}.
\newblock {\emph{\JournalTitle{Sci. Rep.}}} \textbf{\bibinfo{volume}{6}},
  \bibinfo{pages}{29101} (\bibinfo{year}{2016}).

\bibitem{Nemeth_PRL_2008}
\bibinfo{author}{N\'emeth, K.} \emph{et~al.}
\newblock \bibinfo{journal}{\bibinfo{title}{Laser-driven coherent betatron
  oscillation in a laser-wakefield cavity}}.
\newblock {\emph{\JournalTitle{Phys. Rev. Lett.}}}
  \textbf{\bibinfo{volume}{100}}, \bibinfo{pages}{095002}
  (\bibinfo{year}{2008}).

\bibitem{Albert_PRL_2017}
\bibinfo{author}{Albert, F.} \emph{et~al.}
\newblock \bibinfo{journal}{\bibinfo{title}{Observation of betatron x-ray
  radiation in a self-modulated laser wakefield accelerator driven with
  picosecond laser pulses}}.
\newblock {\emph{\JournalTitle{Phys. Rev. Lett.}}}
  \textbf{\bibinfo{volume}{118}}, \bibinfo{pages}{134801}
  (\bibinfo{year}{2017}).

\bibitem{Ferri_PRSTAB_2016}
\bibinfo{author}{Ferri, J.}, \bibinfo{author}{Davoine, X.},
  \bibinfo{author}{Kalmykov, S.~Y.} \& \bibinfo{author}{Lifschitz, A.}
\newblock \bibinfo{journal}{\bibinfo{title}{Electron acceleration and
  generation of high-brilliance x-ray radiation in kilojoule, subpicosecond
  laser-plasma interactions}}.
\newblock {\emph{\JournalTitle{Phys. Rev. Accel. Beams}}}
  \textbf{\bibinfo{volume}{19}}, \bibinfo{pages}{101301}
  (\bibinfo{year}{2016}).

\bibitem{Brabetz_PoP_2015}
\bibinfo{author}{Brabetz, C.} \emph{et~al.}
\newblock \bibinfo{journal}{\bibinfo{title}{Laser-driven ion acceleration with
  hollow laser beams}}.
\newblock {\emph{\JournalTitle{Phys. Plasmas}}} \textbf{\bibinfo{volume}{22}},
  \bibinfo{pages}{013105} (\bibinfo{year}{2015}).

\bibitem{Padgett_OptExpress_2017}
\bibinfo{author}{Padgett, M.~J.}
\newblock \bibinfo{journal}{\bibinfo{title}{Orbital angular momentum 25 years
  on}}.
\newblock {\emph{\JournalTitle{Opt. Express}}} \textbf{\bibinfo{volume}{25}},
  \bibinfo{pages}{11265--11274} (\bibinfo{year}{2017}).

\bibitem{Yao_AdvOptPhoton_2011}
\bibinfo{author}{Yao, A.~M.} \& \bibinfo{author}{Padgett, M.~J.}
\newblock \bibinfo{journal}{\bibinfo{title}{Orbital angular momentum: origins,
  behavior and applications}}.
\newblock {\emph{\JournalTitle{Adv. Opt. Photon.}}}
  \textbf{\bibinfo{volume}{3}}, \bibinfo{pages}{161--204}
  (\bibinfo{year}{2011}).

\bibitem{Allen_PRA_1992}
\bibinfo{author}{Allen, L.}, \bibinfo{author}{Beijersbergen, M.},
  \bibinfo{author}{Spreeuw, R.} \& \bibinfo{author}{Woerdman, J.}
\newblock \bibinfo{journal}{\bibinfo{title}{Orbital angular momentum of light
  and the transformation of laguerre-gaussian laser modes}}.
\newblock {\emph{\JournalTitle{Phys. Rev. A}}} \textbf{\bibinfo{volume}{45}},
  \bibinfo{pages}{8185} (\bibinfo{year}{1992}).

\bibitem{Vieira_PRL_2014}
\bibinfo{author}{Vieira, J.} \& \bibinfo{author}{Mendon\c{c}a, J.~T.}
\newblock \bibinfo{journal}{\bibinfo{title}{Nonlinear laser driven donut
  wakefields for positron and electron acceleration}}.
\newblock {\emph{\JournalTitle{Phys. Rev. Lett.}}}
  \textbf{\bibinfo{volume}{112}}, \bibinfo{pages}{215001}
  (\bibinfo{year}{2014}).

\bibitem{Jorge_arXiv_2018}
\bibinfo{author}{Vieira, J.}, \bibinfo{author}{Mendon\c{c}a, J.~T.} \&
  \bibinfo{author}{Qu\'er\'e}.
\newblock \bibinfo{journal}{\bibinfo{title}{Optical control of the topology of
  laser-plasma accelerators}}.
\newblock {\emph{\JournalTitle{ArXiv e-prints}}}  (\bibinfo{year}{2018}).
\newblock \eprint{1807.04147}.

\bibitem{Cormier_Michel_PhysRevSTAB_2011}
\bibinfo{author}{Cormier-Michel, E.} \emph{et~al.}
\newblock \bibinfo{journal}{\bibinfo{title}{Control of focusing fields in
  laser-plasma accelerators using higher-order modes}}.
\newblock {\emph{\JournalTitle{Phys. Rev. Spec. Top.-Ac.}}}
  \textbf{\bibinfo{volume}{14}}, \bibinfo{pages}{031303}
  (\bibinfo{year}{2011}).

\bibitem{Wang_SciRep_2017}
\bibinfo{author}{Wang, J.~W.}, \bibinfo{author}{Schroeder, C.~B.},
  \bibinfo{author}{Li, R.}, \bibinfo{author}{Zepf, M.} \&
  \bibinfo{author}{Rykovanov, S.~G.}
\newblock \bibinfo{journal}{\bibinfo{title}{Plasma channel undulator excited by
  high-order laser modes}}.
\newblock {\emph{\JournalTitle{Sci Rep.}}} \textbf{\bibinfo{volume}{7}},
  \bibinfo{pages}{16884} (\bibinfo{year}{2017}).

\bibitem{Yao_AdvOptPhot_2011}
\bibinfo{author}{Yao, A.~M.} \& \bibinfo{author}{Padgett, M.~J.}
\newblock \bibinfo{journal}{\bibinfo{title}{Orbital angular momentum: origins,
  behavior and applications}}.
\newblock {\emph{\JournalTitle{Adv. Opt. Photonics}}}
  \textbf{\bibinfo{volume}{3}}, \bibinfo{pages}{161--204}
  (\bibinfo{year}{2011}).

\bibitem{Pariente_OptLett_2015}
\bibinfo{author}{Pariente, G.} \& \bibinfo{author}{Qu\'er\'e, F.}
\newblock \bibinfo{journal}{\bibinfo{title}{Spatio-temporal light springs:
  extended encoding of orbital angular momentum in ultrashort pulses}}.
\newblock {\emph{\JournalTitle{Opt. Lett.}}} \textbf{\bibinfo{volume}{40}},
  \bibinfo{pages}{2037} (\bibinfo{year}{2015}).

\bibitem{Galvez_ProcSPIE_2006}
\bibinfo{author}{Galvez, E.~J.}, \bibinfo{author}{Smiley, N.} \&
  \bibinfo{author}{Fernandes, N.}
\newblock \bibinfo{title}{Composite optical vortices formed by collinear
  laguerre-gauss beams}.
\newblock In \emph{\bibinfo{booktitle}{Proc. SPIE}}, vol.
  \bibinfo{volume}{6131}, \bibinfo{pages}{613105} (\bibinfo{year}{2006}).

\bibitem{Baumann_OptExp_2009}
\bibinfo{author}{Baumann, S.}, \bibinfo{author}{Kalb, D.},
  \bibinfo{author}{MacMillan, L.} \& \bibinfo{author}{Galvez, E.}
\newblock \bibinfo{journal}{\bibinfo{title}{Propagation dynamics of optical
  vortices due to gouy phase}}.
\newblock {\emph{\JournalTitle{Opt. Express}}} \textbf{\bibinfo{volume}{17}},
  \bibinfo{pages}{9818} (\bibinfo{year}{2009}).

\bibitem{Vieira_APS_2017}
\bibinfo{author}{Vieira, J.}
\newblock \bibinfo{title}{{Particle acceleration and exotic light emission in
  structured plasma wakefields}}.
\newblock In \emph{\bibinfo{booktitle}{APS Meeting Abstracts}},
  \bibinfo{pages}{YI3.001} (\bibinfo{year}{2017}).

\bibitem{osiris1}
\bibinfo{author}{Fonseca, R.~A.} \emph{et~al.}
\newblock \bibinfo{title}{Osiris: A three-dimensional, fully relativistic
  particle in cell code for modeling plasma based accelerators}.
\newblock In \bibinfo{editor}{Sloot, P. M.~A.}, \bibinfo{editor}{Hoekstra,
  A.~G.}, \bibinfo{editor}{Tan, C. J.~K.} \& \bibinfo{editor}{Dongarra, J.~J.}
  (eds.) \emph{\bibinfo{booktitle}{Computational Science - ICCS 2002}},
  \bibinfo{pages}{342} (\bibinfo{publisher}{Springer Berlin Heidelberg},
  \bibinfo{address}{Berlin, Heidelberg}, \bibinfo{year}{2002}).

\bibitem{osiris2}
\bibinfo{author}{Fonseca, R.~A.} \emph{et~al.}
\newblock \bibinfo{journal}{\bibinfo{title}{Exploiting multi-scale parallelism
  for large scale numerical modelling of laser wakefield accelerators}}.
\newblock {\emph{\JournalTitle{Plasma Phys. Contr. F.}}}
  \textbf{\bibinfo{volume}{55}}, \bibinfo{pages}{124011}
  (\bibinfo{year}{2013}).

\bibitem{Pak_PRL_2010}
\bibinfo{author}{Pak, A.} \emph{et~al.}
\newblock \bibinfo{journal}{\bibinfo{title}{Injection and trapping of
  tunnel-ionized electrons into laser-produced wakes}}.
\newblock {\emph{\JournalTitle{Phys. Rev. Lett.}}}
  \textbf{\bibinfo{volume}{104}}, \bibinfo{pages}{025003}
  (\bibinfo{year}{2010}).

\bibitem{McGuffey_PRL_2010}
\bibinfo{author}{McGuffey, C.} \emph{et~al.}
\newblock \bibinfo{journal}{\bibinfo{title}{Ionization induced trapping in a
  laser wakefield accelerator}}.
\newblock {\emph{\JournalTitle{Phys. Rev. Lett.}}}
  \textbf{\bibinfo{volume}{104}}, \bibinfo{pages}{025004}
  (\bibinfo{year}{2010}).

\bibitem{Jackson}
\bibinfo{author}{Jackson, J.}
\newblock \emph{\bibinfo{title}{Classical Electrodynamics}}
  (\bibinfo{publisher}{Wiley}, \bibinfo{year}{1998}), \bibinfo{edition}{3rd
  edition} edn.

\bibitem{jrad}
\bibinfo{author}{Martins, J.}, \bibinfo{author}{Martins, S.},
  \bibinfo{author}{Fonseca, R.} \& \bibinfo{author}{Silva, L.}
\newblock \bibinfo{title}{Radiation post-processing in pic codes}.
\newblock In \emph{\bibinfo{booktitle}{Harnessing Relativistic Plasma Waves as
  Novel Radiation Sources from Terahertz to X-Rays and Beyond}}, vol.
  \bibinfo{volume}{7359}, \bibinfo{pages}{73590V} (\bibinfo{publisher}{SPIE},
  \bibinfo{address}{Prague, Czech Republic}, \bibinfo{year}{2009}).

\bibitem{Thomas_PoP_2010}
\bibinfo{author}{Thomas, A.}
\newblock \bibinfo{journal}{\bibinfo{title}{Scalings for radiation from plasma
  bubbles}}.
\newblock {\emph{\JournalTitle{Phys. Plasmas}}} \textbf{\bibinfo{volume}{17}},
  \bibinfo{pages}{056708} (\bibinfo{year}{2010}).

\bibitem{Fourmaux_NJP_2011}
\bibinfo{author}{Fourmaux, S.} \emph{et~al.}
\newblock \bibinfo{journal}{\bibinfo{title}{Demonstration of the
  synchrotron-type spectrum of laser-produced betatron radiation}}.
\newblock {\emph{\JournalTitle{New J. Phys.}}} \textbf{\bibinfo{volume}{13}},
  \bibinfo{pages}{033017} (\bibinfo{year}{2011}).

\bibitem{Glinec_EurPhysLett_2008}
\bibinfo{author}{Glinec, Y.} \emph{et~al.}
\newblock \bibinfo{journal}{\bibinfo{title}{Direct observation of betatron
  oscillations in a laser-plasma electron accelerator}}.
\newblock {\emph{\JournalTitle{Europhys Lett.}}} \textbf{\bibinfo{volume}{81}},
  \bibinfo{pages}{64001} (\bibinfo{year}{2008}).

\bibitem{ADK_JETP_1986}
\bibinfo{author}{Ammosov, M.}, \bibinfo{author}{Delone, N.} \&
  \bibinfo{author}{Krainov, V.}
\newblock \bibinfo{journal}{\bibinfo{title}{Tunnel ionization of complex atoms
  and of atomic ions in an alternating electromagnetic field}}.
\newblock {\emph{\JournalTitle{Sov. Phys. JETP}}}
  \textbf{\bibinfo{volume}{64}}, \bibinfo{pages}{1191} (\bibinfo{year}{1986}).

\end{thebibliography}
\end{document}